\documentclass[12pt]{article}

\usepackage{graphicx} 
\usepackage{amsmath,amssymb}
\usepackage{slashed}
\usepackage{xcolor}
\usepackage{hyperref}
\usepackage{psfrag}

\usepackage{cite}

\newcommand{\veck}{{\bf k}}
\newcommand{\veco}{{\bf 0}}
\newcommand{\veckone}{{\bf k}_1}
\newcommand{\vecktwo}{{\bf k}_2}

\newcommand{\veckj}{{\bf k}_{J}}
\newcommand{\veckji}{{\bf k}_{J,i}}
\newcommand{\veckjone}{{\bf k}_{J,1}}
\newcommand{\veckjtwo}{{\bf k}_{J,2}}
\newcommand{\kjone}{k_{J,1}}
\newcommand{\kjminone}{k_{J {\rm min 1}}}

\newcommand{\kjtwo}{k_{J,2}}
\newcommand{\kjmintwo}{k_{J {\rm min 2}}}

\newcommand{\vecq}{{\bf k}-{\bf k}'}
\newcommand{\vecl}{{\bf l}}
\newcommand{\deins}[1]{{\rm d}#1\,}
\newcommand{\dzwei}[1]{{\rm d}^2#1\,}
\newcommand{\dk}{\dzwei{\veck}}
\newcommand{\dl}{\dzwei{\vecl}}
\newcommand{\dkprime}{\dzwei{\veck'}}
\newcommand{\dkone}{\dzwei{\veckone}}

\newcommand{\dsigma}{\deins{\sigma}}

\newcommand{\dnu}{\deins{\nu}}
\newcommand{\dz}{\deins{z}}
\newcommand{\dx}{\deins{x}}
\newcommand{\dxone}{\deins{x_1}}

\newcommand{\dyjetone}{\deins{y_{J,1}}}
\newcommand{\dyjettwo}{\deins{y_{J,2}}}
\newcommand{\dphij}{\deins{\phi_{J}}}
\newcommand{\dphijone}{\deins{\phi_{J,1}}}

\newcommand{\non}{\nonumber\\}

\newcommand{\chihat}{{\omega}}
\newcommand{\sqrts}{{\sqrt{s}}}

\newcommand{\avgcosm}{\langle \cos m \varphi \rangle}

\begin{document}
\begin{titlepage}

\begin{flushright}
\begin{tabular}{l}
  LPT-ORSAY-14-61 
\end{tabular}
\end{flushright}
\vspace{1.5cm}

\begin{center}
{\LARGE \bf Violation of energy-momentum conservation in Mueller-Navelet jets production}
% {\LARGE \bf Effects of energy non conservation in Mueller-Navelet jets production}
\vspace{1cm}

\renewcommand{\thefootnote}{\alph{footnote}}

{\sc B.~Duclou\'e}${}^{1}$,
{\sc L.~Szymanowski}${}^{2}$,
{\sc S.~Wallon}${}^{1,3}$\footnote{e-mail: \texttt{wallon@th.u-psud.fr}}.
\\[0.5cm]
\vspace*{0.1cm} ${}^1${\it LPT, Universit{\'e} Paris-Sud, CNRS, 91405, Orsay, France} \\[0.2cm]
\vspace*{0.1cm} ${}^2${\it National Centre for Nuclear Research (NCBJ), Warsaw, Poland} \\[0.2cm]
\vspace*{0.1cm} ${}^3${\it UPMC Univ. Paris 06, Facult\'e de Physique, 4 place Jussieu, 75252 Paris Cedex 05, France} \\[1.0cm]
{\it \large
%\today
 }
\vskip2cm
{\bf Abstract:\\[10pt]} \parbox[t]{\textwidth}{
We study effects related to violation of energy-momentum conservation inherent to the BFKL approach, in the particular case of Mueller-Navelet jets production. We argue, based on the comparison of the lowest order non trivial corrections $\mathcal{O}(\alpha_s^3)$ to the  cross section with predictions of an exact calculation, that the inclusion of next-to-leading order BFKL corrections to the jet production vertex significantly reduces the importance of these effects.}
\vskip1cm
\end{center}

\vspace*{1cm}
\end{titlepage}

\section{Introduction}

Many processes have been proposed as a way to probe the high energy dynamics of
QCD, described by the Balitsky-Fadin-Kuraev-Lipatov (BFKL)
approach~\cite{Fadin:1975cb,Kuraev:1976ge,Kuraev:1977fs,Balitsky:1978ic}.
A general weakness of this approach is the fact that it does not respect strict energy-momentum conservation. While such kinematic constraints are in principle subleading in the BFKL approach, numerically their effect could be sizable.
There have been many attempts to estimate these effects of energy-momentum non-conservation, for example in refs.~\cite{DelDuca:1994ng,Orr:1997im,Kwiecinski:1999yx}. 

A phenomenological way to take these effects into account was proposed in ref.~\cite{DelDuca:1994ng}. The authors studied dijet production at large rapidity intervals and compared the results of an exact $\mathcal{O}(\alpha_s^3)$ contribution with the ones obtained in the leading logarithmic (LL) BFKL framework. It was found that a LL BFKL calculation strongly overestimates the cross section with respect to an exact treatment.

One can avoid this issue by using a numerical method based on a Monte Carlo event generator, which iterates over the number of emitted gluons. It is then possible to impose energy-momentum conservation at each iteration. This approach was followed by the authors of ref.~\cite{Orr:1997im}, where it was confirmed that this effect is significant.

In ref.~\cite{Kwiecinski:1999yx}, it was shown that imposing consistent kinematical constraint within the leading order BFKL Green's function can lead to corrections equivalent to about 75\% of effects generated by the NLO corrections to the BFKL kernel.

A point of special interest is to study this violation of energy-momentum conservation in the production of forward jets separated by a large interval of rapidity $Y$ at hadron colliders, called Mueller-Navelet jets~\cite{Mueller:1986ey}. This process was proposed as a promising observable which permits to reveal effects of BFKL dynamics.
The authors of ref.~\cite{Kwiecinski:2001nh} followed the proposal of ref.~\cite{DelDuca:1994ng} based on the introduction of an effective rapidity interval $Y_{\rm eff}$ to study energy-momentum conservation effects in this process. The outcome of this work is that taking this effect into account in a LL framework leads to a much better description of Tevatron data on the azimuthal correlations of these jets.
In the same spirit, a study with LO vertices and NLL Green's function was performed in ref.~\cite{Marquet:2007xx}.

Recently we performed a comprehensive study of Mueller-Navelet jets production within a full NLL BFKL framework at the LHC~\cite{Ducloue:2013hia,Ducloue:2013bva}. It is natural to expect that after taking into account NLL BFKL corrections the effects due to non-conservation of energy-momentum should be less severe than at LL accuracy. The aim of the present paper is to quantify the correctness of this expectation by extending the method proposed in ref.~\cite{DelDuca:1994ng} beyond the leading logarithmic accuracy.

The content of this paper is the following: in section 2 we summarize shortly, based on ref.~\cite{DelDuca:1994ng}, the problem of non conservation of energy-momentum in the context of Mueller-Navelet jets production at LL accuracy. In section 3, we show how still staying at the level of $\mathcal{O}(\alpha_s^3)$, this problem can be mostly cured by including the NLO corrections to the jet production vertex~\cite{Ciafaloni:1998kx,Ciafaloni:1998hu,Bartels:2001ge,Bartels:2002yj,Caporale:2011cc,Hentschinski:2011tz,Chachamis:2012cc}.

\section{Effect of non conservation of energy-momentum at LL}

We proceed in a close analogy with the idea proposed in ref.~\cite{DelDuca:1994ng} but adopting the same conventions as in refs.~\cite{Ducloue:2013hia,Ducloue:2013bva}. Thus we will study the angular coefficients $\mathcal{C}_m$ defined as
\begin{equation}
  \mathcal{C}_m = \int d\varphi \cos{(m \varphi)} \frac{\dsigma}{{\rm d}|\veckjone|\,{\rm d}|\veckjtwo|\,\dyjetone \dyjettwo d\varphi} \,,
\end{equation}
where $\veckjone$ and $\veckjtwo$ are the transverse momenta of the jets, $y_{J,1}$ and $y_{J,2}$ their rapidities ($Y=|y_{J,1}-y_{J,2}|$) and $\varphi$ is the relative azimuthal angle $\varphi=\pi-\phi_{J,1}-\phi_{J,2}$ (for more details we refer to refs.~\cite{Ducloue:2013hia,Ducloue:2013bva}). In the following we will also use the notation $k_{J,i}\equiv|\veckji|$. Let us emphasize that for $m=0$ we recover the cross section, while values of $m$ different from $0$ give access to the azimuthal correlations according to $\mathcal{C}_m=\avgcosm$.

\begin{figure}[h]
  \psfrag{in1}{}
  \psfrag{in2}{}
  \psfrag{out1}{$y_1$}
  \psfrag{out2}{$y_2$}
  \psfrag{out3}{$y_3$}
  \begin{minipage}{0.49\textwidth}
    \psfrag{gap1}{}
    \psfrag{gap2}{}
    \centering\includegraphics[height=4cm]{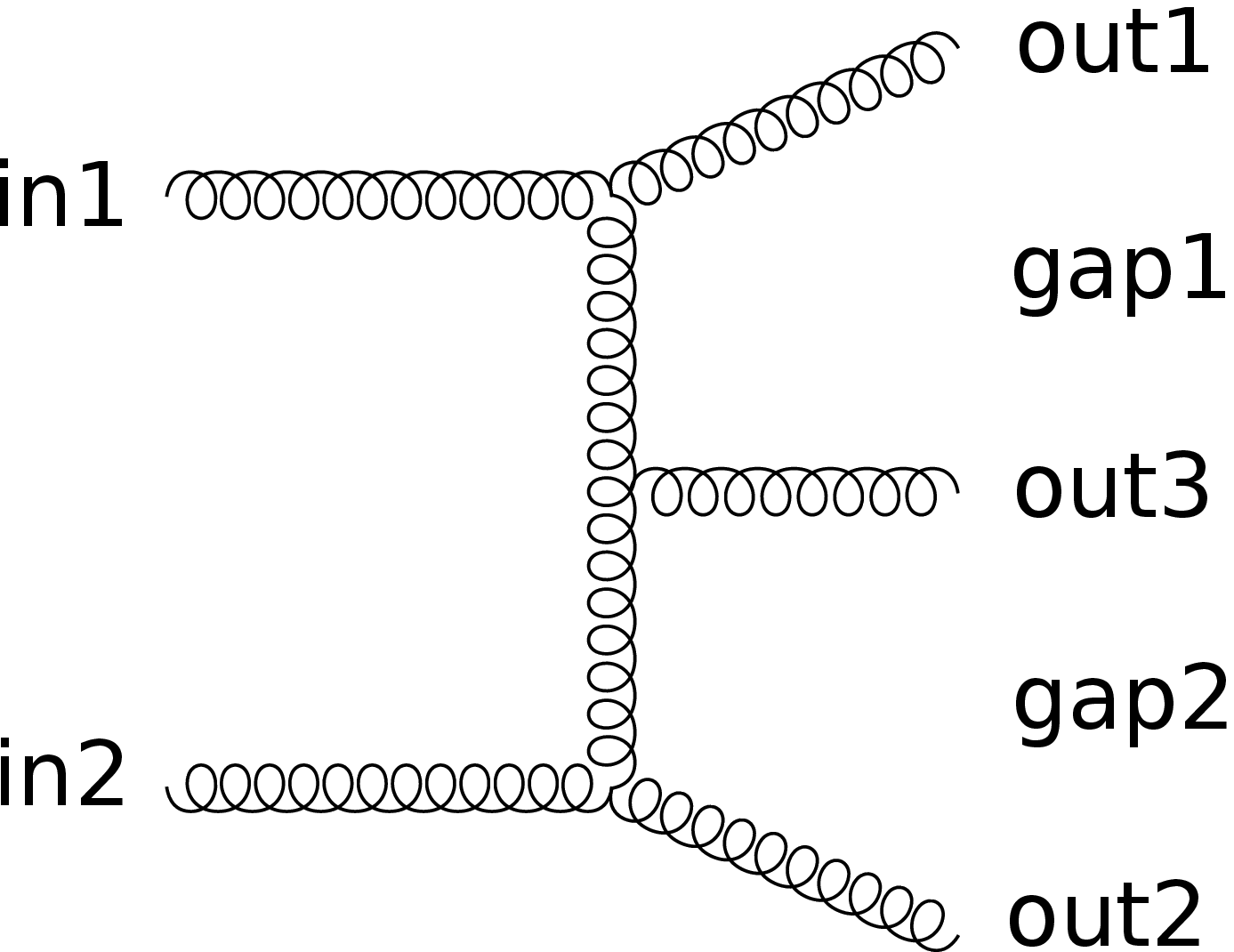}
  \end{minipage}
  \begin{minipage}{0.49\textwidth}
    \psfrag{gap1}{\scalebox{0.8}{\hspace{-1cm}large rapidity gap}}
    \psfrag{gap2}{\scalebox{0.8}{\hspace{-1cm}large rapidity gap}}
    \centering\includegraphics[height=4cm]{figures/2to3.eps}
  \end{minipage}
  \caption{Schematic representation of the $gg\to ggg$ process in an exact calculation (L) and a LL BFKL treatment (R).}
  \label{Fig:two-to-three_exact_LL}
\end{figure}

The main idea of the approach of ref.~\cite{DelDuca:1994ng} is to study the exact $\mathcal{O}(\alpha_s^3)$ contribution to $\mathcal{C}_m$, $\mathcal{C}_m^{2\to3}$, corresponding to the fusion of two incoming partons into three outgoing partons, as shown schematically on fig.~\ref{Fig:two-to-three_exact_LL} (L). This exact result is then compared with the one obtained in the BFKL approximation, $\mathcal{C}_m^{{\rm BFKL},\mathcal{O}(\alpha_s^3)}$, by expanding the LL BFKL result in powers of $\alpha_s$ and truncating to order $\mathcal{O}(\alpha_s^3)$.
This lead the authors of ref.~\cite{DelDuca:1994ng} to define an effective rapidity $Y_{\rm eff}$ as
\begin{equation}
  Y_{\rm eff} \equiv\ Y \frac{\mathcal{C}_m^{2\to3}}{\mathcal{C}_m^{{\rm BFKL},\mathcal{O}(\alpha_s^3)}} \,.
  \label{eq:Cm_e-m_cons}
\end{equation}
The definition of the effective rapidity~(\ref{eq:Cm_e-m_cons}) is motivated by the observation that if one replaces $Y$ by $Y_{\rm eff}$ in the BFKL calculation, expands in powers of $\alpha_s$ and truncates to order $\alpha_s^3\,,$ the exact result is recovered.
Thus the use of $Y_{\rm eff}$ instead of $Y$ in the BFKL expression can correct in an effective way the potentially too strong assumptions made in a BFKL calculation while preserving the additional emissions of gluons specific to this approach.
The value of $Y_{\rm eff}$ is an indication of how valid the BFKL approximation is: a value close to $Y$ means that this approximation is valid, whereas a value significantly different from $Y$ means that it is a too strong assumption in the kinematics under study.

Below, for simplicity, we will only consider the case of incoming gluons, so that we restrict ourselves to the $gg \to ggg$ subprocess\footnote{In any case, extending the analysis we will present here to take into account quark contributions would not present any conceptual difficulty.}. In ref.~\cite{DelDuca:1994ng} the authors also took into account the quark contributions, which turn out to have a very small influence on the value of $Y_{\rm eff}$ for large rapidity difference.
We have checked that considering only gluonic contributions we reproduce the results of ref.~\cite{DelDuca:1994ng} with good accuracy.

Let us recall the general expression of $\mathcal{C}_m$ in the BFKL approach 
\begin{align}
  \mathcal{C}_m & = (4-3\delta_{m,0}) \int \dnu C_{m,\nu}(|\veckjone|,x_{J,1})C^*_{m,\nu}(|\veckjtwo|,x_{J,2})e^{\chihat(m,\nu)Y} \,,
\end{align}
with
\begin{equation}
C_{m,\nu}(|\veckj|,x_{J})
= \int\dphij\dk \dx f(x) V(\veck,x)E_{m,\nu}(\veck)\cos(m\phi_J) \,,
\end{equation}
where
\begin{equation}
  E_{m,\nu}(\veck) = \frac{1}{\pi\sqrt{2}}\left(\veck^2\right)^{i\nu-\frac{1}{2}}e^{im\phi}\,,
\label{def:eigenfunction}
\end{equation}
$\phi$ being the azimuthal angle of $\veck$. At LL accuracy, the jet vertex $V$ reads
\begin{equation}
  V^{(0)}(\veck,x) = h^{(0)}(\veck)\,\mathcal{S}_J^{(2)}(\veck;x) \,,
  \label{eq:vertex_lo}
\end{equation}
where
\begin{equation}
  h^{(0)}(\veck) = \frac{\alpha_s}{\sqrt{2}}\frac{C_{A}}{\veck^2} \,,
\end{equation}
with $C_A=N_c=3$, and
\begin{equation}
  \mathcal{S}_J^{(2)}(\veck;x) = \delta\left(1-\frac{x_J}{x}\right)|\veckj|\delta^{(2)}(\veck-\veckj) \,.
  \label{def:sj2}
\end{equation}
The LL BFKL eigenvalue is
\begin{equation}
  \omega(m,\nu)=\bar{\alpha}_s\chi_0(m,\nu) \,,
\end{equation}
with $\bar{\alpha}_s=\alpha_s\frac{N_c}{\pi}$, and
\begin{equation}
\chi_0(m,\nu) = 2\Psi(1)-\Psi\left(\frac{1+|m|}{2}+i\nu\right)-\Psi\left(\frac{1+|m|}{2}-i\nu\right)\,,
\label{def:chi_0}
\end{equation}
where $\Psi(x) = \Gamma'(x) /\Gamma(x)$.

Using these formulas, the coefficients $\mathcal{C}_m$ up to leading logarithmic accuracy have the form
\begin{equation}
  \mathcal{C}_m = \frac{\left(\alpha_s C_{A}\right)^2}{\veckjone^2 \veckjtwo^2}
  x_{J,1}f(x_{J,1}) x_{J,2}f(x_{J,2})
  \int \dnu \left(\frac{\veckjone^2}{\veckjtwo^2}\right)^{i\nu} e^{\bar{\alpha}_s\chi_0(m,\nu)Y} \,.
  \label{def:cm}
\end{equation}
Eq.~(\ref{def:cm}) can be expanded in powers of $\alpha_s$ as
\begin{equation}
   \mathcal{C}_m = \frac{\left(\alpha_s C_{A}\right)^2}{\veckjone^2 \veckjtwo^2}
  x_{J,1}f(x_{J,1}) x_{J,2}f(x_{J,2})
  \int \dnu \left(\frac{\veckjone^2}{\veckjtwo^2}\right)^{i\nu} \left( 1+\bar{\alpha}_s\chi_0(m,\nu)Y+\dots  \right) \,.
  \label{eq:Cm_LL_expanded}
\end{equation}
The only $\mathcal{O}(\alpha_s^3)$ term contributing to the denominator of eq.~(\ref{eq:Cm_e-m_cons}) comes from the second term of the expansion of the Green's function and reads
\begin{equation}
  \mathcal{C}_m^{{\rm BFKL},\mathcal{O}(\alpha_s^3)}=\frac{\left(\alpha_s C_{A}\right)^2}{\veckjone^2 \veckjtwo^2}
  x_{J,1}f(x_{J,1}) x_{J,2}f(x_{J,2})
  \int \dnu \left(\frac{\veckjone^2}{\veckjtwo^2}\right)^{i\nu} \bar{\alpha}_s\chi_0(m,\nu)Y \,.
  \label{eq:Cm_LL_alphas3}
\end{equation}
It corresponds to the case where only one gluon emission is taken into account in the Green's function, as shown schematically on fig.~\ref{Fig:two-to-three_exact_LL} (R). On this figure we can see 
the main source of discrepancy between the exact result and the BFKL one at order $\mathcal{O}(\alpha_s^3)$: if we denote the rapidities of the most forward and most backward final-state partons by $y_1$ and $y_2$ respectively, in the exact treatment the rapidity $y_3$ of the third parton can lie anywhere between $y_1$ and $y_2$. On the contrary, it is assumed in the BFKL calculation that there is a strong ordering in rapidity, i.e. we have $y_2 \ll y_3 \ll y_1$.
In the exact calculation, the longitudinal momentum fractions of the incoming partons, $x_a$ and $x_b$, depend on the kinematics of the three outgoing partons according to
\begin{equation}
  x_a = \frac{k_1 \, e^{y_1} + k_2  \, e^{y_2} + k_3 \,  e^{y_3}}{\sqrt{s}} \:,\:\;\:\:\:\: x_b = \frac{k_1  \, e^{-y_1} + k_2  \, e^{-y_2} + k_3  \, e^{-y_3}}{\sqrt{s}} \,,
  \label{eq:x_exact}
\end{equation}
where $k_i$ is the transverse momentum of outgoing parton $i$.
In this case the integration over $x_a$ and $x_b$ is not trivial since these variables depend on $y_3$. When integrating over $y_3$, the configurations where $y_3$ is close to the borders of the domain of integration, i.e. close to $y_1$ or $y_2$, are strongly suppressed by the parton distribution functions.
This does not happen in the LL BFKL approach, as in this case the longitudinal momentum fractions of the incoming partons are taken equal to the ones of the jets according to
\begin{equation}
  x_a = x_{J,1} = \frac{k_1  \, e^{y_1}}{\sqrt{s}} \:,\:\;\:\:\:\:
  x_b = x_{J,2} = \frac{k_2  \, e^{-y_2}}{\sqrt{s}} \,,
  \label{eq:x_approx}
\end{equation}
which is the limit of eq.~(\ref{eq:x_exact}) in the case $y_2 \ll y_3 \ll y_1$.
These values do not depend on $y_3$, this is why we could factor out the parton distribution functions in eq.~(\ref{eq:Cm_LL_expanded}). This approximation means that the PDFs are probed at values which do not depend on the value of $k_3$ and $y_3$. As a consequence, the suppression effect for $y_3$ close to $y_1$ or $y_2$ present in the exact calculation is neglected. The integration over $y_3$ is reduced to a global factor $|y_1-y_2|=Y$, as seen in eq.~(\ref{eq:Cm_LL_alphas3}).

\begin{figure}[h]
  \psfrag{Tevatron}[l][l][.8]{$\sqrt{s}=1.8$ TeV, $Y=6$}
  \psfrag{LHC}[l][l][.8]{$\sqrt{s}=7$ TeV, $Y=8$}
  \psfrag{yeffy}[l][l][.8]{$Y_{\rm eff}/Y$}
  \psfrag{kj1}[l][l][1]{\hspace{0.1cm}$\kjtwo$ (GeV)}
  \centering\includegraphics[width=10cm]{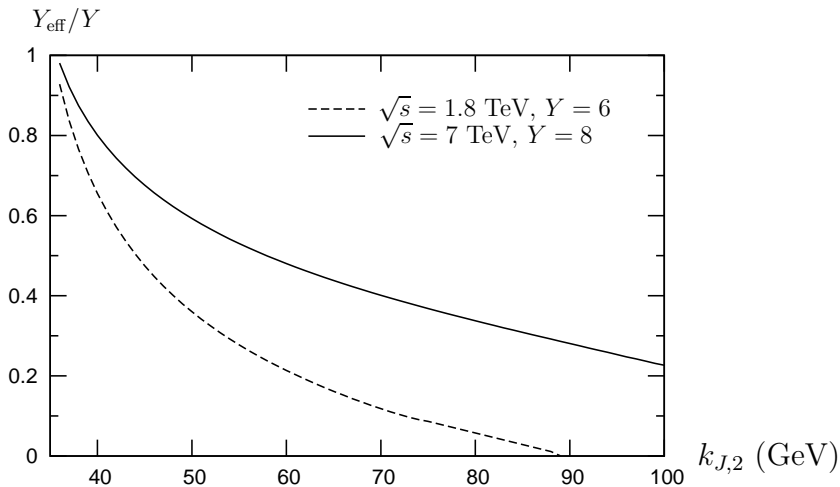}
  \caption{Variation of $Y_{\rm eff}/Y$ as defined in eq.~(\ref{eq:Cm_e-m_cons}) as a function of $\kjtwo$ at fixed $\kjone=35$ GeV in two kinematic configurations: $Y=6$ at $\sqrt{s}=1.8$ TeV (dashed line) and $Y=8$ at $\sqrt{s}=7$ TeV (solid line).}
  \label{Fig:yeff_LO}
\end{figure}

When $\kjone \ne \kjtwo$ it is possible to perform the integration over $\nu$ in eq.~(\ref{eq:Cm_LL_alphas3}) analytically by using the integral representation of the $\psi$ function
\begin{equation}
\psi(z) = \int_0^1 dx \frac{1-x^{z-1}}{1-x} - \gamma \,,
\label{psif}
\end{equation}
where $\gamma$ is the the Euler constant $\gamma \approx 0.577215$. We get
\begin{equation}
  \mathcal{C}_m^{{\rm BFKL},\mathcal{O}(\alpha_s^3)}=(-1)^m \frac{\alpha_s^3 C_{A}^2}{\kjone\kjtwo} N_c
  \left( \frac{\kjtwo}{\kjone} \right)^m \frac{x_{J,1} \, f(x_{J,1})  \, x_{J,2} \, f(x_{J,2})}{|\kjone-\kjtwo|(\kjone+\kjtwo)}\,Y \,,
\end{equation}
to be compared with $\mathcal{C}_m^{2\to3}$, obtained in an exact calculation at order $\mathcal{O}(\alpha_s^3)$~\cite{Gottschalk:1979wq,Kunszt:1979ci,Berends:1981rb,Parke:1986gb}. The ratio $Y_{\rm eff}/Y$ is shown for fixed $\kjone=35$ GeV as a function of $\kjtwo$ on fig.~\ref{Fig:yeff_LO}, for $m=0$ which corresponds to the cross-section, for kinematics typical of the Tevatron ($\sqrt{s}=1.8$ TeV, $Y=6$) and of the LHC ($\sqrt{s}=7$ TeV, $Y=8$). One can see that while the effective rapidity is close to $Y$ when the two jets have similar transverse momenta, the ratio $Y_{\rm eff}/Y$ decreases quickly when $\kjtwo$ increases, indicating that the LL BFKL calculation overestimates the cross section by a large amount. We also observe that, as expected, this effect is less severe in the LHC kinematics than in the Tevatron ones since the larger center of mass energy makes the high energy limit more justified.
Nevertheless, this effect is still sizable as long as the transverse momenta of the jets are not very close to each other.
This observation is important since in refs.~\cite{Ducloue:2013hia,Ducloue:2013bva} we compared our results with the ones obtained by a fixed order calculation in an asymmetric configuration ($\kjminone \ne \kjmintwo$), necessary to obtain trustable results in the fixed order approach.
Therefore one could be worried that our results are not reliable because of this energy-momentum non conservation issue. However, one could expect that going to higher orders, such as NLL, would make this issue less problematic since these corrections take into account some effects which were neglected at LL accuracy.

\section{Next-to-leading order}

In this section we will show how the fast dropping of the ratio $Y_{\rm eff}/Y$ with increasing $|\kjone-\kjtwo|$ can be avoided to a large extent by the inclusion of the NLO corrections to the jet vertex when evaluating $Y_{\rm eff}$ as defined in eq.~(\ref{eq:Cm_e-m_cons}).

Let us remind that the NLL corrections to Mueller-Navelet jets production come from two sources: the Green's function and the jet vertex (for detailed formulas we refer to refs.~\cite{Ducloue:2013hia,Ducloue:2013bva}).
The NLL corrections to the Green's function are beyond the $\mathcal{O}(\alpha_s^3)$ precision we are interested in.
Indeed, in this case, the expansion of the Green's function $e^{\omega Y}$ reads $1+\bar{\alpha}_s \chi_0(m,\nu) Y+\bar{\alpha}_s^2 \chi_1(m,\nu) Y+\dots$ which, taking into account the global $\alpha_s^2$ factor coming from the two jet vertices, means that the NLL corrections to the Green's function play no role at order $\alpha_s^3$.
On the contrary, the NLO corrections to the jet vertices, giving rise to an extra power of $\alpha_s$, contribute at order $\mathcal{O}(\alpha_s^3)$ when convoluted with the first term of the expansion of the Green's function.

A major difference between the LO and the NLO jet vertex is the fact that at next-to-leading order there can be two emitted partons instead of one at LO. These two partons are not separated by a large interval of rapidity (the NLO corrections to the jet vertices give rise to an extra power of $\alpha_s$ without an extra power of $\ln{\hat s}$). Therefore, when considering the BFKL result truncated at $\mathcal{O}(\alpha_s)^3$, contributions where two of the three partons are not separated by a large rapidity gap appear. These kind of contributions, which are present in the exact $2\to3$ calculation, were neglected in the previous section. Therefore one can expect that taking into account these contributions leads to results closer to the exact ones.
This is especially important since in ref.~\cite{DelDuca:1994ng}, the authors observed that the large overestimate of the cross section at LL comes mostly from using the approximate values of $x$'s in the PDFs given by eq.~(\ref{eq:x_approx}) instead of the exact ones (\ref{eq:x_exact}), thus neglecting the strong suppression of configurations with $y_3$ close to $y_1$ or $y_2$. On the contrary, with the NLO jet vertex, the longitudinal momentum fraction of an incoming parton is no longer fixed to be equal to the one of the corresponding outgoing jet, making it necessary to perform the integration over $x_1$ and $x_2$ numerically.

\begin{figure}[h]
  \psfrag{in1}{}
  \psfrag{in2}{}
  \psfrag{out1}{$y_1$}
  \psfrag{out2}{$y_2$}
  \psfrag{out3}{$y_3$}
  \psfrag{gap1}{\scalebox{0.8}{\hspace{-1cm}large rapidity gap}}
  \psfrag{gap2}{\scalebox{0.8}{\hspace{-1cm}large rapidity gap}}
  \begin{minipage}{0.49\textwidth}
    \centering\includegraphics[height=4cm]{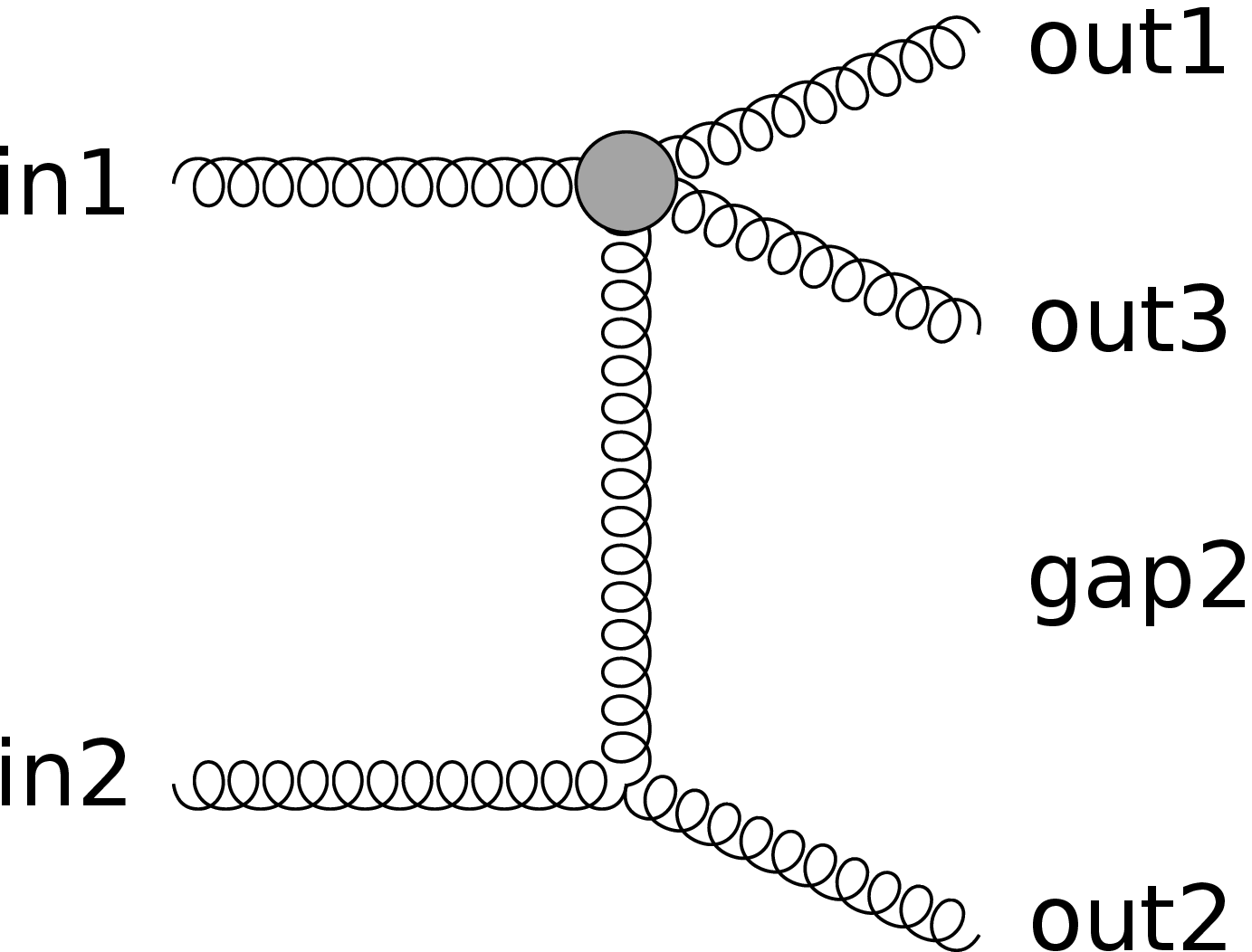}
  \end{minipage}
  \begin{minipage}{0.49\textwidth}
    \centering\includegraphics[height=4cm]{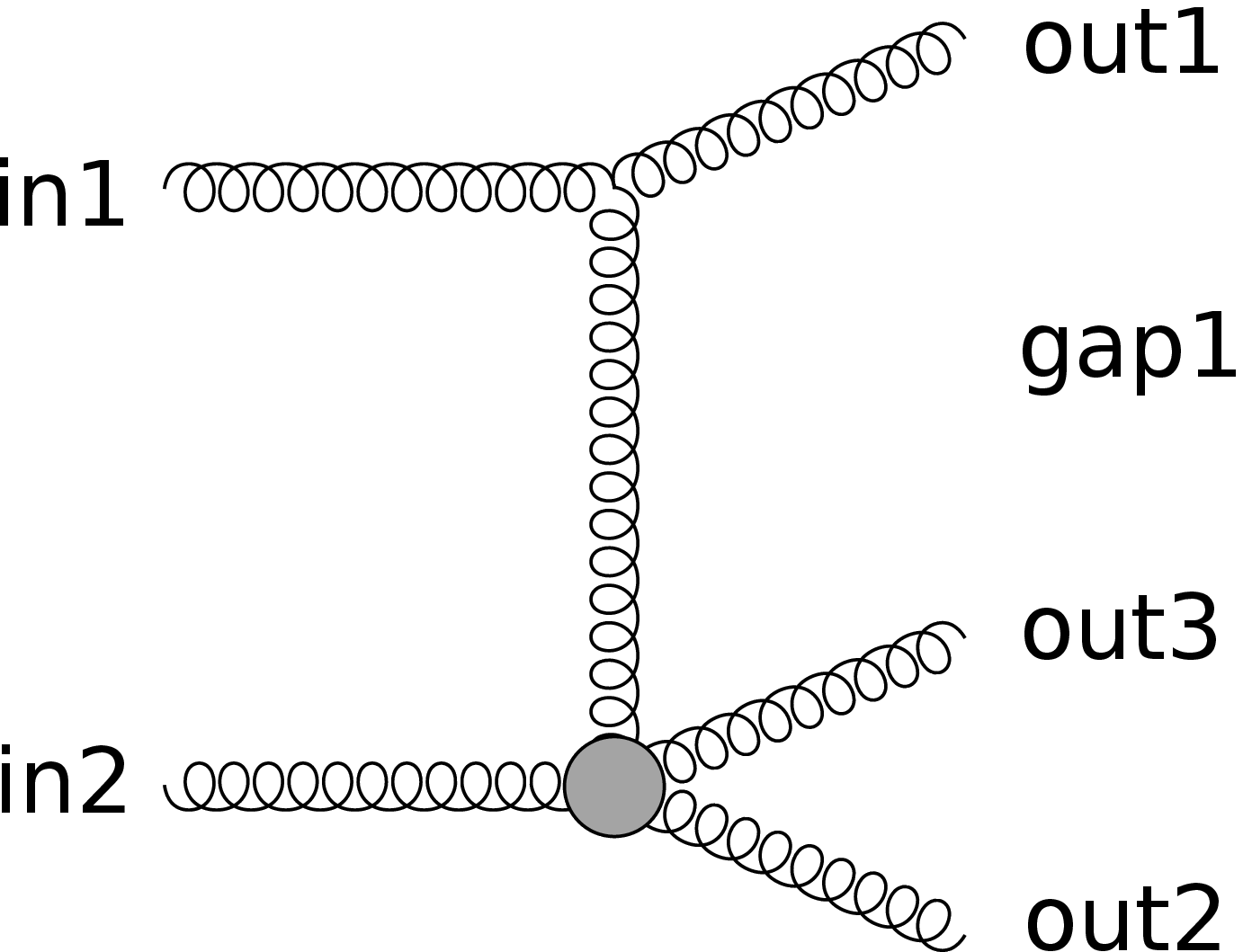}
  \end{minipage}
  \caption{Additional contributions to the $2 \to 3$ process coming from the NLO jet vertex in a NLL BFKL calculation.}
  \label{Fig:two-to-three_NLO_vertex}
\end{figure}

The two additional terms that we need to include when considering the jet vertex at next-to-leading order are illustrated on fig.~\ref{Fig:two-to-three_NLO_vertex}. When compared with fig.~\ref{Fig:two-to-three_exact_LL} (R), they lead to additional contributions coming from the jet vertices: at NLO, two partons can be produced with no large rapidity separation. One then recovers some of the contributions of fig.~\ref{Fig:two-to-three_exact_LL} (L) that were missing in fig.~\ref{Fig:two-to-three_exact_LL} (R), where $y_3$ is similar to $y_1$ or $y_2$.

\begin{figure}[h]
  \psfrag{p1}{\raisebox{-.2cm}{\footnotesize $\hspace{-.1cm}\veco, x$}}
    \psfrag{a}{\footnotesize $\hspace{-.1cm}\veck$}
  \hspace{0cm}\begin{minipage}{0.3\textwidth}
    \psfrag{d}{\footnotesize $\veck, x$}
    \centering\includegraphics[height=3cm]{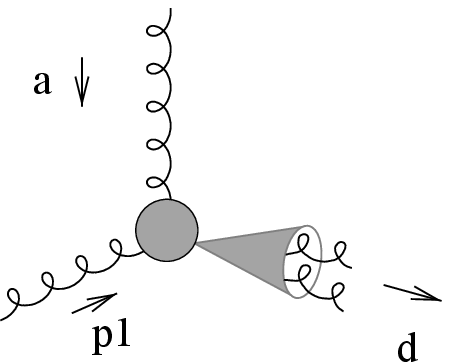}
    
    \vspace{0.2cm}
    
    {\small $\mathcal{S}_J^{(2)}(\veck,x)$}
    
    \vspace{0.2cm}
    
    (a)
  \end{minipage}
  \begin{minipage}{0.3\textwidth}
    \psfrag{c}{\footnotesize $\vecq, x \, z$}
    \psfrag{d}{\footnotesize $\veck', x(1-z)$}
    \centering\includegraphics[height=3cm]{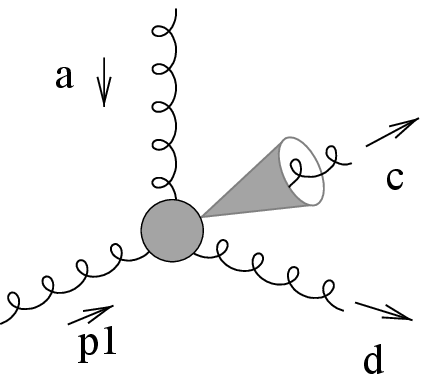}
    
    \vspace{0.2cm}
    
    {\small $\mathcal{S}_J^{(2)}(\vecq,xz)$}
    
    \vspace{0.2cm}
    
    (b)
  \end{minipage}
  \hspace{0.7cm}\begin{minipage}{0.3\textwidth}
    \psfrag{c}{\footnotesize $\vecq, x \, z$}
    \psfrag{d}{\footnotesize $\veck', x(1-z)$}
    \centering\includegraphics[height=3cm]{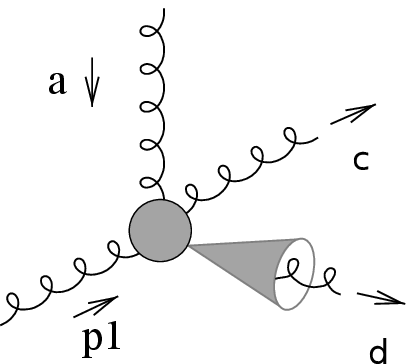}
    
    \vspace{0.2cm}
    
    {\small $\mathcal{S}_J^{(2)}(\veck',x(1-z))$}
    
    \vspace{0.2cm}
    
    (c)
  \end{minipage}
  \caption{Contributions to the real emission of the NLO jet vertex.}
  \label{Fig:jet_vertex_real}
\end{figure}

To include these two contributions, we start again from the expression of the coefficients $\mathcal{C}_m$ used before,
\begin{align}
  \mathcal{C}_m & = (4-3\delta_{m,0}) \int \dnu C_{m,\nu}(|\veckjone|,x_{J,1}) \, C^*_{m,\nu}(|\veckjtwo|,x_{J,2}) \, e^{\chihat(m,\nu)Y} \non
  & =  (4-3\delta_{m,0}) \int \dnu C_{m,\nu}(|\veckjone|,x_{J,1}) \, C^*_{m,\nu}(|\veckjtwo|,x_{J,2}) \left( 1+\chihat(m,\nu)Y+\dots \right) \,.
\end{align}
The expression for the jet vertex to be used in $C_{m,\nu}(|\veckj|,x_{J})$ is now
\begin{equation}
  V(\veck,x)=V^{(0)}(\veck,x)+\alpha_s V^{(1)}(\veck,x) \,,
\end{equation}
where $V^{(1)}(\veck,x)$ are the NLO corrections which can be read from ref.~\cite{Colferai:2010wu}.
In the expression of $V^{(1)}(\veck,x)$ an important quantity is the function $\mathcal{S}_J^{(3)}$ which determines how, in the case of real corrections, one should deal with the two outgoing partons: if the two partons are emitted 'close' to each other, they should be combined and form the jet. Otherwise, one should sum the two contributions corresponding to the case where the jet is constituted by either of these two partons. These three possibilities are shown on fig.~\ref{Fig:jet_vertex_real}.
The exact form of $\mathcal{S}_J^{(3)}$ depends on the practical jet algorithm that is used for the calculation (which determines the condition to consider that two partons are 'close' enough to each other to be combined into a jet).
It is a sum of three contributions, shown on fig.~\ref{Fig:jet_vertex_real}, which involve different arguments of $\mathcal{S}_J^{(2)}$, as defined in eq.~(\ref{def:sj2}).
In this work we will use the cone algorithm\footnote{Note that the difference between the cone and anti-$k_t$ algorithms is small in our NLL BFKL treatment of Mueller-Navelet jets~\cite{Ducloue:2013bva}.} with a size $R_{{\rm cone}}=0.5$.

The $\mathcal{O}(\alpha_s^3)$ contribution to $\mathcal{C}_m$ is
\begin{equation}
  \mathcal{C}_m^{{\rm BFKL},\mathcal{O}(\alpha_s^3)}=\mathcal{C}_m^{{\rm BFKL},\mathcal{O}(\alpha_s^3) \, {\rm LL}} + \mathcal{C}_m^{{\rm BFKL},\mathcal{O}(\alpha_s^3) \, {\rm NLL}}
\end{equation}
where $\mathcal{C}_m^{{\rm BFKL},\mathcal{O}(\alpha_s^3) \, {\rm LL}}$ was already calculated~(\ref{eq:Cm_LL_alphas3}) and the NLL contribution reads
\begin{align}
  \mathcal{C}_m^{{\rm BFKL},\mathcal{O}(\alpha_s^3) \, {\rm NLL}}=(4-3\delta_{m,0}) \Bigg( & \int \dnu C^{{\rm NLO}}_{m,\nu}(|\veckjone|,x_{J,1}) \, C^{* {\rm LO}}_{m,\nu}(|\veckjtwo|,x_{J,2}) \non
  & + \int \dnu C^{{\rm LO}}_{m,\nu}(|\veckjone|,x_{J,1}) \, C^{* {\rm NLO}}_{m,\nu}(|\veckjtwo|,x_{J,2}) \Bigg)
  \label{eq:Cm_NLL_alphas3}
\end{align}
with
\begin{align}
  C^{{\rm LO}}_{m,\nu}(|\veckj|,x_J) & =\int\dphij\dk \dx f(x) V^{(0)}(\veck,x)E_{m,\nu}(\veck)\cos(m\phi_J) \non
  & = \frac{\alpha_s C_A}{2}\left(\veckj^2\right)^{i\nu-1}x_Jf(x_J)(1+\delta_{m,0})
\end{align}
and
\begin{equation}
  C^{{\rm NLO}}_{m,\nu}(|\veckj|,x_J) =\alpha_s \int\dphij \dk \dx f(x) V^{(1)}(\veck,x)E_{m,\nu}(\veck)\cos(m\phi_J) \,.
\end{equation}
Let us now focus on the first term of eq.~(\ref{eq:Cm_NLL_alphas3})
\begin{equation}
  \mathcal{C}_m^{{\rm BFKL},\mathcal{O}(\alpha_s^3) \, {\rm NLL} (1)} = (4-3\delta_{m,0}) \int \dnu C^{{\rm NLO}}_{m,\nu}(|\veckjone|,x_{J,1}) \, C^{* {\rm LO}}_{m,\nu}(|\veckjtwo|,x_{J,2}) \,.
\end{equation}
It can be written as
\begin{align}
  \mathcal{C}_m^{{\rm BFKL},\mathcal{O}(\alpha_s^3) \, {\rm NLL} (1)} = & (4-3\delta_{m,0})\,\alpha_s \int \dnu \frac{\alpha_s C_A}{2}\left(\veckjtwo^2\right)^{-i\nu-1}x_{J,2}f(x_{J,2})(1+\delta_{m,0}) \non
  & \times \int \dphijone \dkone \dxone f(x_1) V^{(1)}(\veckone,x_1)E_{m,\nu}(\veckone)\cos(m\phi_{J,1}) \non
  = & (2-\delta_{m,0})\,\frac{\alpha_s^2 C_A}{\veckjtwo^2} x_{J,2}f(x_{J,2}) \non
  & \times \int \dnu \dphijone \dkone \dxone \left(\veckjtwo^2\right)^{-i\nu} f(x_1) V^{(1)}(\veckone,x_1)E_{m,\nu}(\veckone)\cos(m\phi_{J,1}) \non
  = & \frac{2-\delta_{m,0}}{\pi\sqrt{2}}\,\frac{\alpha_s^2 C_A}{\veckjtwo^2} x_{J,2}f(x_{J,2}) \int \dnu \dphijone \dkone \dxone \left(\frac{\veckone^2}{\veckjtwo^2}\right)^{i\nu} \non
  & \times f(x_1) V^{(1)}(\veckone,x_1)\frac{1}{|\veckone|}e^{i m \phi_1}\cos(m\phi_{J,1})
\end{align}
where in the last step we have used the explicit representation of the LL BFKL eigenfunctions (\ref{def:eigenfunction}).
The integration over $\nu$ gives
\begin{equation}
  \int d\nu \left( \frac{\veckone^2}{\veckjtwo^2} \right)^{i \nu}=\pi |\veckjtwo| \delta(|\veckone|-|\veckjtwo|)
\end{equation}
so that
\begin{align}
  \mathcal{C}_m^{{\rm BFKL},\mathcal{O}(\alpha_s^3) \, {\rm NLL} (1)} = & \frac{2-\delta_{m,0}}{\pi\sqrt{2}}\,\frac{\alpha_s^2 C_A}{\veckjtwo^2} x_{J,2}f(x_{J,2}) \int \dphijone \dkone \dxone \pi |\veckjtwo| \delta(|\veckone|-|\veckjtwo|) \non
  & \hspace{4.5cm} \times f(x_1) V^{(1)}(\veckone,x_1)\frac{1}{|\veckone|}e^{i m \phi_1}\cos(m\phi_{J,1}) \non
  = & \frac{2-\delta_{m,0}}{\sqrt{2}}\,\frac{\alpha_s^2 C_A}{\kjtwo} \, x_{J,2}f(x_{J,2}) \non
  & \times \int \dphijone  \deins{\phi_1} \dxone f(x_1) V^{(1)}(\kjtwo,\phi_1,x_1) e^{i m \phi_1}\cos(m\phi_{J,1})
  \label{eq:Cm_NLL_alphas3_final}
\end{align}
where $V^{(1)}(\kjtwo,\phi_1,x_1)$ is to be understood as $V^{(1)}(\veckone,x_1)$ where $|\veckone|=\kjtwo$. Therefore only three integrations remain, over $\phi_1$, $\phi_{J,1}$ and $x_1$. The second term of eq.~(\ref{eq:Cm_NLL_alphas3}), $\mathcal{C}_m^{{\rm BFKL},\mathcal{O}(\alpha_s^3) \, {\rm NLL} (2)}$, is obtained in the same way by exchanging jets $1$ and $2$.

As we are interested in the case $\kjone \ne \kjtwo$, the expression of $V^{(1)}$ entering in eq.~(\ref{eq:Cm_NLL_alphas3_final}) can be significantly simplified since all terms proportional to $\delta(k_1-\kjone)$ vanish: those containing $\mathcal{S}_J^{(2)}(\veck,\cdots)$ or $V^{(0)}(\veck,\cdots)$ with $\veck=\veckone$.
After integrating over $k_1$ using the procedure above, these terms will be proportional to $\delta(\kjtwo-\kjone)$, so they vanish in the case $\kjone \ne \kjtwo$.
Thus in this case the expression for $V^{(1)}$ is given by\footnote{Since we are only considering contributions from gluons, we put $N_f=0$.}:
\begin{eqnarray}
&&V^{(1)}(\veck,x)\non
 &=& \frac{C_A}{\pi}\int_0^1 \frac{\dz}{1-z} \left[(1-z)P(1-z)\right]\int\frac{\dl}{\pi\vecl^2}\non
&&\hspace{1cm}\times\frac{\mathcal{N}C_A}{\vecl^2+(\vecl-\veck)^2}\Big[
\mathcal{S}_J^{(3)}(z\veck+(1-z)\vecl,(1-z)(\veck-\vecl),x(1-z);x)\non
&&\hspace{1cm}\hphantom{\times\frac{\mathcal{N}C_A}{\vecl^2+(\vecl-\veck)^2}\Big[}
+\mathcal{S}_J^{(3)}(\veck-(1-z)\vecl,(1-z)\vecl,x(1-z);x)\Big]\non
&&\hspace{-.2cm}+\frac{C_A}{\pi}\int\frac{\dkprime}{\pi}\int_0^1\dz\Bigg[ P(z)
(1-z)\frac{(\vecq)\cdot\big((1-z)\veck-\veck'\big)}{(\vecq)^2\big((1-z)\veck-\veck'\big)^2} h_{\rm g}^{(0)}(\veck')\non
&&\hspace{-.2cm}\hphantom{+\frac{C_A}{\pi}\int\frac{\dkprime}{\pi}\int_0^1\dz\Bigg[}\times\mathcal{S}_J^{(3)}(\veck',\vecq,xz;x)
\non
&&\hspace{-.2cm}\hphantom{+\frac{C_A}{\pi}\int\frac{\dkprime}{\pi}\int_0^1\dz\Bigg[ }
-\frac{1}{z(\vecq)^2}\Theta\big(|\vecq|-z(|\vecq|+|\veck'|)\big)V_{\rm g}^{(0)}(\veck',x)\Bigg]\,.\,\,
\end{eqnarray}
A further simplification of this expression comes from the fact that the first term in the expression of $\mathcal{S}_J^{(3)}$ corresponding to fig.~\ref{Fig:jet_vertex_real} (a) vanishes.
This is due to the fact that the process is initiated by collinear partons, i.e. with no transverse momentum. Since there is no transverse momentum in the initial state, the same is true for the final state and so $\veckone+\vecktwo+{{\bf k}_3}=\veco$.
Considering fig.~\ref{Fig:two-to-three_NLO_vertex} (L) as an example, we see that if partons $1$ and $3$ are to be combined into a jet, we have $\veckjone=\veckone+{{\bf k}_3}=-\vecktwo=-\veckjtwo$ (since the lower vertex is treated at leading order), and so $\kjone=\kjtwo$. The opposite is true: if we impose $\kjone \ne \kjtwo$, partons $1$ and $3$ can't form a single jet.

To quantify the influence of the terms (\ref{eq:Cm_NLL_alphas3}) on the value of the effective rapidity $Y_{\rm eff}$, 
we will only consider the case $m=0$ corresponding to the cross section, but this analysis could also be in straightforward way performed for the azimuthal correlations.

\begin{figure}[h]
  \psfrag{LO}[l][l][.8]{LL at $\mathcal{O}(\alpha_s^3)$}
  \psfrag{NLO}[l][l][.8]{NLL at $\mathcal{O}(\alpha_s^3)$}
  \psfrag{yeffy}[l][l][.8]{$Y_{\rm eff}/Y$}
  \psfrag{kj1}[l][l][1]{\hspace{0.1cm}$\kjtwo$ (GeV)}
  \centering\includegraphics[width=10cm]{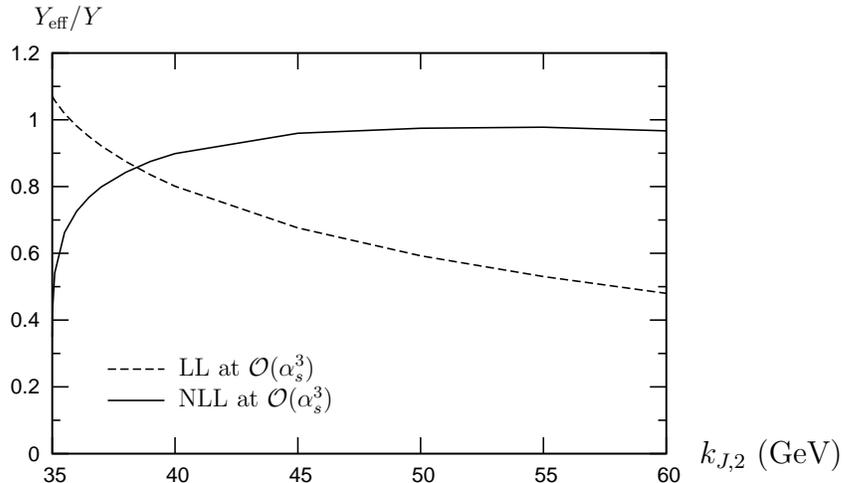}
  \caption{Variation of $Y_{\rm eff}/Y$ as defined in eq.~(\ref{eq:Cm_e-m_cons}) as a function of $\kjtwo$ at fixed $\kjone=35$ GeV for $Y=8$ and $\sqrts=7$ TeV at leading logarithmic (dashed) and next-to-leading logarithmic (solid) accuracy.}
  \label{Fig:yeff_NLO}
\end{figure}

On fig.~\ref{Fig:yeff_NLO} we show the ratio $Y_{\rm eff}/Y$ for fixed $\kjone=35$ GeV as a function of $\kjtwo$ at a center of mass energy of $7$ TeV, both in the LL approximation and NLL approximation.
As we have seen in the previous section, in the LL case this ratio decreases quickly with increasing $\kjtwo$. The behavior is different when including NLO corrections to the jet vertex, as the ratio first grows and then stabilizes close to $1$ for $\kjtwo \gtrsim 45$ GeV.
The dip when $\kjone$ is close to $\kjtwo$ is probably due to the fact that, even if we have removed several contributions explicitly proportional to $\delta(\kjone-\kjtwo)$, some additional contributions divergent when $\kjone \to \kjtwo$ may appear when performing the integrations numerically. A more careful analysis would be needed to isolate such terms but the analytical study of the NLL amplitude is much more complicated than at LL accuracy.
The fact that this dip is smeared around $\kjone=\kjtwo$ is probably due to the fact that we are performing a numerical treatment.
Nevertheless, we would like to stress that our calculation should be trustable in the region where the transverse momenta of the jets are significantly different. In this region, the value of $Y_{\rm eff}$ is very close to $Y$ and this value is very stable with respect to $\kjtwo$. 
This means that in this region the inclusion of the NLO corrections to the jet vertex dramatically reduces the overestimate of the cross section found in a LL calculation.
Such a reduction is important for trustable comparison of predictions obtained within BFKL approach with the ones from fixed order calculations which are reliable only in asymmetric configurations.

\section{Conclusions}

In this paper, we have studied the 
importance of violation of energy-momentum conservation  
in Mueller-Navelet jets production in the BFKL NLL approach. This is an important question in the context of LHC measurements, with the aim of getting a clear signal of high-energy resummation effects.
We have shown, based on the study of the $2 g \to 3 g$ process,
at order $\mathcal{O}(\alpha_s^3),$ treated either exactly or based on a NLL BFKL approximation, that when including NLO vertex corrections which means here allowing the third gluon to be close in rapidity with respect to the two most forward gluons (which is not allowed in the LL BFKL approximation), one obtains a very significant improvement of energy-momentum conservation. This is true in the region where the two outgoing jets have not very similar  transverse momenta, which is the region of main interest in view of comparisons with fixed order NLO computations which suffer from instabilities when the two jet transverse momenta are almost identical. We thus believe that energy-momentum non conservation in NLL BFKL should not be a major issue in future phenomenological studies.

\section*{Acknowledgments}

We thank Cyrille Marquet and Christophe Royon for discussions.

This work is supported by the French Grant PEPS-PTI, the
Polish Grant NCN No.~DEC-2011/01/B/ST2/03915 and  the Joint Research Activity Study of Strongly Interacting Matter (HadronPhysics3, Grant Agreement n.283286) under the 7th Framework Programme of the European Community.

\end{document}